\documentstyle[psfig]{l-aa}

\begin{document}  \thesaurus{ 11.01.2 ; 11.09.1 NGC 1068 ; 11.14.1 ; 11.19.1}

\title{  Near-IR  Images  of the Torus and  Micro-Spiral  Structure  in
NGC 1068  using  Adaptive  Optics
\thanks{ Based on observations obtained at Canada-France-Hawaii Telescope
operated by the National Research Council of Canada, The Centre National de la
Recherche Scientifique de France and the University of Hawaii } } \author{ D.
Rouan \inst{1} \and F.  Rigaut \inst{2} \and D.  Alloin \inst{3} \and R.  Doyon
\inst{4} \and O.  Lai \inst{1} \and D.  Crampton \inst{5} \and E.  Gendron
\inst{1} \and R.  Arsenault \inst{2,1} }
 \offprints{ D.  Rouan (rouan@obspm.fr) }

\institute{  Observatoire de  Paris --  D\'epartement  Spatial -- CNRS
URA 264 -- F92195  Meudon  Cedex, France
\and  Canada-France-Hawaii  Telescope  corp., PO Box 1597, Kamuela,
Hawaii, 96743, USA
 \and  Service  d'Astrophysique -- CNRS URA 2052 -- CE Saclay, F91191
 Gif sur Yvette,  France
 \and  Universit\'e de Montr\'eal - D\'epartement  d'Astronomie -- C.P.
 6128, Succ. A, Montr\'eal, H3C 3J7, Canada
 \and Dominion Astrophysical Observatory, HIA,  National Research Council of Canada, RR5 Victoria, V8X 4M6, Canada}

   \date{ received...   ; accepted ...  }

     \maketitle

\markboth{ TORUS AND MICRO-SPIRAL STRUCTURE IN
NGC 1068} { TORUS AND MICRO-SPIRAL STRUCTURE IN
NGC 1068}

\begin{abstract}

We present  diffraction-limited  near-IR  images in J, H and K of the nucleus of
NGC 1068,  obtained  with the  Adaptive  Optics  system {\it Pueo} at CFHT.  The
achieved  resolution   (0\farcs12)  reveals  several  components,   particularly
prominent on the [J$-$K] image:  a) an unresolved,  conspicuous core (size $<$ 9
pc); b) an elongated structure at P.A.  $\approx$ 102\degr, beginning to show up
at radius $\approx$ 15 pc; c) a S-shaped  structure with radial extent $\approx$
20 pc, including a bar-like central  elongation at P.A.  $\approx$  15\degr\ and
two short  spiral arms.  A precise  registration  of the IR peak was carried out
relative to the HST I-band peak.  The K unresolved  core is found to be close to
the location of the putative  central  engine  (radio  source S1).  Consistent
with the Unified  Model of AGN, the near-IR core is likely the emission from the
hot inner walls of the  dust/molecular  torus.  The  extremely red colors of the
0\farcs2 diameter core, [J-K]=7.0,  [H-K]=3.8, lead to an intrinsic  extinction
A$_V \geq $25, assuming classical dust grains at 1500 K.

The elongated  structure at P.A.  $\approx$  102\degr\ may trace the presence of
cooler dust within and around the torus.  This  interpretation  is  supported by
two facts at least:  a) the elongated  structure is  perpendicular  to the local
radio jet originating at S1; b) its direction  follows  exactly that of the disk
of ionized gas recently  found with the VLBA.  Regarding  the S-shaped  feature,
the near-IR flux of the bar-like central elongation at P.A.= $\approx$ 4\degr\ ,
if  interpreted  in terms of free-free  emission  from  ionized  gas, is roughly
consistent  with the  level  of 5 GHz  emission.  However,  the  radio  spectrum
behaviour is indicative of synchrotron  emission and we rather interpret the 2.2
$\mu $m emission as originating  from warm dust in the shaded part of NLR clouds
or in stellar  photospheres.  The shape  itself  suggests an  extremely  compact
barred  spiral  structure,  that  would be the  innermost  of a series of nested
spiral structures, as predicted by models and simulations.  This is supported by
the inner  stellar  distribution  -- deduced  from the J image -- which  clearly
follows an exponential disk with a 19 pc  scale-length,  precisely that expected
from the rotation of a bar twice this size.


\end{abstract}

\section{Introduction}

Considerable  effort has recently been directed towards high spatial  resolution
imaging of the nucleus of NGC 1068 from UV to radio  wavelengths (UV \& visible:
Macchetto  et al.  1994;  Capetti  et al.  1995,  1997;  near-IR:  Chelli  et al.
1987;  Gallais, 1991; Young et al.  1996; Marco et al.  1997; mid-IR:  Braatz et
al.  1993;  Cameron et al.  1993; radio:  Wilson \& Ulvestad  1987;  Planesas et
al.  1991;  Blietz  et al.  1994,  Tacconi  et al.  1994;  Muxlow  et al.  1996;
Gallimore  et al.  1996a,b,  1997 ; Greenhill  \& Gwinn  1996).  Indeed, as the
closest  Seyfert 2 nucleus, NGC 1068  deserves  such  efforts  since it is a key
object in  investigating  models of active  galactic  nuclei (AGN).  The popular
``Unified''  model (e.g., Antonucci 1993) includes a parsec-scale  torus of dust
and  molecular gas around the central  engine.  Models of the infrared  emission
from the torus  (Krolik \& Begelman  1986; Pier \& Krolik  1993;  Efstathiou  \&
Rowan-Robinson  1994; Granato, Danese \& Franceschini  1997) explore torus sizes
from 1 to 100 pc.  Recently,  Gallimore et al.  (1997) detected a 1 pc elongated
distribution of ionized gas at 8~GHz with the VLBA which they interpreted as the
``hot zone" of obscuring  material  surrounding the AGN.  However, direct images
showing the torus, or any elongated  structure, are still lacking in the near-IR
where it is expected to be most conspicuous.  To achieve this goal, one requires
both high angular resolution  (1\arcsec = 72 pc at the distance of NGC 1068) and
high  contrast,  the light from the central  region  being a complex  mixture of
starlight,  synchrotron  emission and excited gas  emission, all  scattered  and
absorbed by a clumpy dust component.

Near-IR  imaging at high  angular  resolution  offers  potential
advantages in the study of AGN because:  {\it i)} the wavelength  range
lies between two domains carrying  complementary pieces of information
--  the visible (excited gas in the NLR), and the thermal IR-radio
range (cool dust and  synchrotron  emission from electrons),   {\it
ii)}  the   extinction   is  reduced,   {\it iii)} diffraction-limited
images are now possible thanks to adaptive optics (AO hereafter).
Here, we present  new results  obtained  with {\sl Pueo}, the AO
system  recently commissioned on the 3.6 m Canada-France-Hawaii
Telescope.

\section{ Observations and results} \label{S-obs}

The CFHT AO system, based on a concept by Roddier et al.  (1991), uses a 19 zone
bimorph  mirror  controlled  by a curvature  wavefront  sensor  (WFS) to produce
diffraction-limited  observations (FWHM  $\sim$0\farcs12) in the near-IR (Rigaut
et al., 1994, Lai et al., 1996).  The MONICA infrared  camera, a facility  instrument of the
Universit\'e  de Montr\'eal  (Nadeau al.  1994), was mounted at the output focus
of {\sl Pueo}, itself installed at the Cassegrain focus of CFHT.  Special optics
in MONICA give a scale on the Nicmos-3 array of 0\farcs0344 per pixel.

The  observations  of NGC 1068, and a nearby blank sky reference  position, were
obtained on Feb 14, 16 and 18 1997.  A dither pattern, with relative  offsets of
$\sim$0\farcs8,  was used in order to reduce  the  effect of bad  pixels  and to
improve the  flat-fielding,  which was  initially  derived from dome flats.  The
UKIRT  faint  standard  stars  FS8 and FS7 were  used to  provide  flux  and PSF
calibrations.  The AO system was servoed on the nucleus  itself; its rather high
brightness  and the  quality  of the  seeing  (ranging  from  0\farcs38  in K to
0\farcs7 in J) allowed good  correction of the  turbulence  and Strehl ratios of
over 60\% were obtained at K.  The true point spread function (PSF) in each band
was recovered using WFS information  (V\'eran et al.  1997).  More specifically,
it can be shown  that  departure  of a long  exposure  PSF from a  perfect  Airy
pattern is contributed by :  {\it i)} the non-corrected  static aberrations, due
for  instance  to  the IR  camera  optics  :  these  can be  measured  using  an
artificial  source ; {\it  ii)} the  non-perfect  compensation  by the  adaptive
mirror of the WF  deformations :  at low spatial  frequencies  they are actually
measured by the WFS but cannot be completely accomodated by the mirror, while at
high spatial frequencies, they are not measured by the WFS, but can be estimated
using a Kolmogorov model of the atmosphere,  scaled on the actual seeing, itself
evaluated from the amplitude of the correction  (V\'eran et al.  1997).  V\'eran
(1997) showed that this method for  retrieving  the  real-time PSF is valid, the
source being point-like or extended.  Similarly, one may wonder if the exactness
of the AO correction is maintained,  using a slightly  extended  reference as in
the case of NGC  1068\footnote{  We have  measured  on the image of NGC 1068, as
seen by the WFS (see Sect 2.3), a FWHM of 0\farcs14  for the  brightest  spot on
which the AO loop is  closed.}.  Indeed, it has been shown  already that the use
of  an  extended   source   instead  of  a   point-like   one  to  perform   the
curvature-sensing  and AO  correction  only  reduces  the  efficiency  of the WF
sensing, but not its accuracy  (Rousset,  1994).  Moreover, the source extension
in the case of NGC 1068 is smaller  than the seeing  value for all  images :  in
such  conditions,  even the  efficiency  loss is small  (Equ.  13 in  Rousset,
1994).  In  conclusion,  the non  point-like  shape of the reference  should not
introduce  systematic  errors in the structure of the AO corrected image, nor on
the  reconstructed  PSF.  This is  illustrated  by the  comparison  between  the
recovered  PSF at K and the image of a nearby  star  (Fig.  1e,  upper  left and
lower left respectively).

Image processing proceeded as follows:  {\it i)} bad pixel correction; {\it ii)}
sky  subtraction,  using a  median-averaged  sky estimate; {\it iii)} flat-field
correction;  {\it  iv)}   re-centering  of  the  different   exposures   through
cross-correlation  techniques;  {\it v)}  adjustment  of the sky level among the
overlapping  regions to produce a homogeneous  background; {\it vi)} co-addition
of the  overlapping  regions,  rejecting  deviant  pixels  (clipped  mean).  The
resulting  images  were then  deconvolved  using the  classical  Lucy-Richardson
algorithm (60 iterations) while following the recipe recently proposed by Magain
et al.  (1997) to constrain the final PSF.

\begin{figure*}[htbp]
\centerline{\psfig{figure=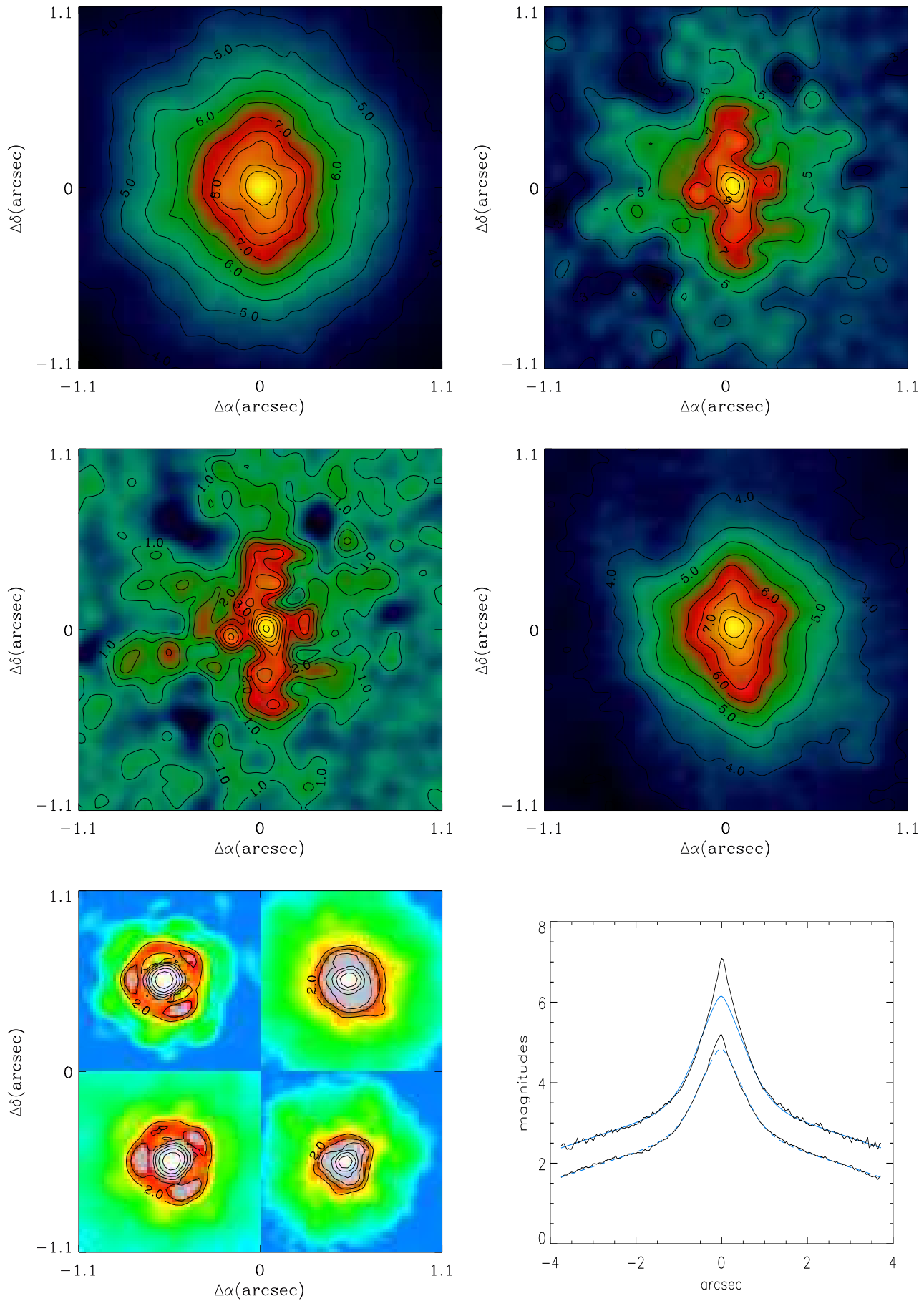,width=17cm}}
\caption{  False color  images and isophotes of the 2.2\arcsec $\times$ 2.2\arcsec\
area around the nucleus of NGC 1068. For all images, the
pixel size is 0\farcs0344 and a log (magnitude)  scale is used : the
step in magnitude between two isophotes is 0.5 for (a),(c), (d) and (e), and
1.0 for (b).   From
the top, left to right:  (a)  un-deconvolved  K ; (b) deconvolved K (see
text) ; (c) deconvolved  [J$-$K] ; (d) deconvolved  H; (e) four panels
showing PSF images  recovered  using the WFS information: upper left --
K, upper right -- H, lower right -- J, lower  left -- the PSF observed
in K on a nearby  star;  (f) normalized  magnitude profiles of a cut at P.A.  =
125$^{\circ}$ in J (lower) and H (upper); a vertical shift has been applied for
clarity. For each profile a fit by 2 exponential disks  (see text)
is also displayed (dashed blue).  }
\label{F-raw}
\end{figure*}

The resulting K image is presented in Fig.  \ref{F-raw}-a,  on a magnitude (log)
scale, chosen because AO provides a high dynamic range  (typically 1.3 10$^4$ at
K) and significant details are seen at all flux levels.  In Fig.  \ref{F-raw}-e,
we show a set of PSF  images:  the  PSFs  recovered  using  the WFS  information
(V\'eran et al.  1997) and used in the deconvolution procedure are shown for the
3 bands and, for  comparison,  the PSF  observed  in K on a nearby  star is also
shown.  We show in Figs.  \ref{F-raw}-b  and -c the H and K deconvolved  images.
Within the 1\arcsec\ inner region the following features are visible, especially
on the deconvolved K image:  \\ {\it a)} An unresolved  core with size less than
0\farcs12,  i.e.  less than 8.6 pc.  The brightest pixel in the core is a factor
$\approx$~1300  above the  background  (measured at 6\arcsec\ away from the core
and  showing  a mean S/N of 10).  This  core was  known  already  ( Marco et al.
1997),  although we can place a more stringent upper limit on its size.  \\ {\it
b)} An  ESE-WNW  elongated  structure  at PA  $\approx$  102$^{\circ}$,  roughly
perpendicular  to the axis of the inner ionizing cone (P.A.  = 15$^{\circ}$,  as
originally  derived by Evans et al., 1991).  This structure starts to show up at
a  radius  0\farcs20  with  a  contrast\footnote{The   contrast  is  defined  as
$(I_{struct}  -  I_{avg})/  I_{avg}$,  i.e.  the ratio of the flux excess in the
structure  at the  relevant  radius,  with  respect to the mean  inter-structure
brightness  measured at the same  radius.}  of 0.28 in the ESE  quadrant  and of
0.19 in the WNW quadrant.  The mean  brightness per pixel of this structure is a
factor 340 above the background.  \\ {\it c)} Along the NS direction, a S-shaped
feature  extends  over  0\farcs3  on each side of the  unresolved  core,  with a
brightness  per  pixel  200 times the  background  level.  The  contrast  of the
structure at a radius of 0\farcs2 is 0.29 to the N and 0.75 to the S.  Moreover,
within the S-shaped  structure,  there is an  elongation in the direction of the
axis of the ionizing cone. We hereafter refer to it as the bar-like structure.

In order to ascertain the reality of the  structures we see at faint
levels, we have built maps of NGC 1068, of the observed PSF (star) and
of the recovered PSF, normalizing  the flux at each  position  ($\rho,
\theta$) to the flux  averaged along an annulus of the same radius
($\rho$).  These maps  reveal, especially  at faint  brightness
levels  (less than 2\% of the peak), a few radial  structures that are
residual  aberrations  not  compensated  by AO (mostly  due to the
camera optics). In particular,  a faint feature
at P.A.  $\approx 50^{\circ}$ is present on all maps that is clearly such an artifact.
However, none of the features in NGC 1068 mentioned above has such a
spurious counterpart in the PSF map, and their brightness is well above
those of the artifacts.

Owing to the complexity of the central region, we have discarded a simple fit to
the stellar  distribution in terms of elliptical  isophotes.  Instead, we assume
that the J band  emission is a  satisfactory  representation  of the  underlying
stellar  component and we have derived [J$-$K] and [J$-$H] color images in order
to minimize effects of the stellar  component.  The [J$-$K]  isophotes (see Fig.
\ref{F-raw}-d) delineate particularly well the three features already mentioned,
i.e., the core, the P.A.  $\approx$  102\degr\  structure  and the  S-shaped  NS
structure.  Similar information is also available at 1.65~$\mu $m in the [J$-$H]
color image (not  shown).  Neither of the two  extended  features (b) and (c) is
apparent on the J image.

\subsection{Photometry of the inner core}

Regarding  the core observed in J, H and K,  integration  of the K flux within a
0\farcs2 diameter circular diaphragm on the  un-deconvolved  image gives K = 9.3
mag.  PSF fitting with FWHM = 0\farcs12  gives a result in excellent  agreement,
confirming  that indeed the core is unresolved  at 2.2 $\mu $m.  In J and H, the
contribution of the underlying  extended component must first be subtracted.  To
accomplish  this, the J and H profiles at P.A.  = 125\degr, a direction  free of
the small scale  structures  discussed  previously,  were extracted.  Beyond r =
0\farcs15,  these profiles are quite well fitted by two  exponential  disks with
characteristic radii of 3\farcs1 and 0\farcs26 (see Fig.  1-f and the discussion
in section 3).  Once the  contribution  of the  extended  component  is removed,
aperture  photometry  of the 0\farcs2  core gives J = 16.3 mag and H = 13.1 mag.
Assuming  that the core centers are  coincident in J, H and K, this leads to the
extremely red colors  [J$-$K] = 7.0 and [H-K] = 3.8.  For  comparison, we find a
mean  value of [J$-$K] = 3.5 for the  regions  situated  at r = 0\farcs2  either
along the P.A.  = 102\degr\ elongated structure or along the bar-like elongation
within the S-shaped feature at P.A.  = 4\degr.

\subsection{Large and small scale spiral  structures}

We compare the inner S-shaped  feature with other  bar/spiral  structures in NGC
1068.  From the archived  F547M  WFPC/HST image in the continuum  around 547 nm,
one can distinguish  ({\it i}) an outer barred spiral structure, with the bar at
P.A.  =  43$^{\circ}$,  extending over 16\arcsec\ in diameter, and ({\it ii}) an
intermediate  barred  spiral  structure,  with the bar at P.A.  =  26$^{\circ}$,
extending over 3\farcs3 in diameter.  Finally, on the current near-IR images, we
find that the inner S-shaped feature shows a bar-like central elongation at P.A.
=  4$^{\circ}$  which  extends  over a  scale  of  0\farcs5.  The  existence  of
interwoven   spiral/barred   structures  in  galaxies  has  been   suggested  by
simulations  (Friedli \& Martinet, 1993; Heller \& Shlosman 1994; Combes 1994a).
They are thought to build a fueling  channel for the active nucleus in AGN.  NGC
1068, in which we detect overlapping spiral structures on three different scales
(1.15 kpc, 240 pc, 36 pc) at a P.A.  rotating from  43$^{\circ}$  to 4$^{\circ}$
inward,  appears to be a good test case for  detailed  modeling of this  effect.
This is deferred to a future paper.

\subsection{Relative  positioning  of the  near-IR  and  visible peaks}

In order to interpret the new data in the perspective of the AGN modeling, it is
necessary to  accurately  register the near-IR peak with  respect to the visible
one.  One way to achieve this is through simultaneous  visible/IR  observations,
as  carried  out by Marco et al.  (1997),  who  found the K peak to be offset by
0\farcs28$\pm$0.05  S and  0\farcs08$\pm$0.05  W from the optical (I band) peak.
Here, we have to rely on  observations of a nearby star interlaced  with the NGC
1068  measurements.  This star provides a positional  reference  both on the WFS
and on the near-IR camera.  Moreover, because the nearby stellar light source is
coincident  in the  visible  and  near-IR,  any  relative  offset in the near-IR
between the nearby stellar peak and the NGC 1068 peak reflects an intrinsic  separation
between the visible and the near-IR sources within NGC 1068.  However, since the
WFS  bandpass is rather wide and the central  region of NGC 1068 quite  complex,
the image of NGC 1068 as seen by the WFS has to be  generated.  Such a composite
image was  synthesized  from  archival  F502N, F547M, F658N and F791N  WFPC2/HST
images, kindly made available to us in their fully reduced and precisely aligned
form by Z.  Tsetanov.  These four images were properly  scaled,  weighted by the
WFS  response  and  summed to provide  the  required  image.  Because  the F547M
intensity  peak is widely used as the  positional  reference  for the  ``visible
peak"  (Lynds et al.  1991) for NGC 1068  research,  we have  compared  the {\it
Pueo} WFS  composite  image, just  obtained  above, to the F547M image.  We find
that the center of gravity of the Pueo WFS image is 0\farcs011 N and  0\farcs007
E of  the  visible  peak.  Finally  we  have  derived  the  offset  between  the
unresolved  core  on the K  image  of  NGC  1068  and  its  visible  peak  to be
0\farcs180$\pm0.030$ S and  0\farcs153$\pm0.030$  W.  This result broadly agrees
with that cited above by Marco et al.  (1997) and that  derived by Thatte et al.
(1997)  (0\farcs216$\pm$0.100  S and  0\farcs095$\pm$  0.100 W, as deduced  from
their Fig.  6, although no indication is given about which image of NGC 1068 was
used for the WFS).

\section{Discussion and conclusion}

Within the remaining  uncertainty  in the location of the K peak in NGC 1068, we
are led to  conclude  that the  unresolved  core in K (size  less  than 8 pc) is
coincident with the radio component S1 (Gallimore et al.  1996a), with the 12.4
$\mu $m  source  observed  by  Braatz  et al.  (1993),  and with the  center  of
symmetry  of the  UV/optical  polarization  map  (Capetti  et  al.  1995).  As a
result,  we  consider  that the K  unresolved  core can be  identified  with the
immediate  surroundings of the central engine, most probably with hot dust close
to sublimation.  Assuming classical dust grains at 1500 K, with associated [J-K]
= 2.76 and [H-K] = 1.18, the extreme red colors of the 0\farcs2  diameter  core,
[J-K]  = 7.0  [H-K]  = 3.8  provide  an  extinction  A$_V$  of 25  and  41  mag,
respectively  (Rieke \& Lebofsky 1985).  The  discrepancy  between the two A$_V$
figures obtained remains to be elucidated, and might be related to the nature of
the dust grains.  We also notice,  given the  0\farcs12  resolution  of the {\sl
Pueo} data set, that there is some evidence from the profile  analysis that the
core is  resolved  in H and J,  while  this is not the  case in K.  This  result
suggests some contribution from scattered light in J and H.

 How could we interpret the two extended structures also detected in the K and H
bands?  The  ESE-WNW  belt at  P.A.=102$^{\circ}$  extends  over  about 20 pc on
either  side of the core.  We argue that it {\it is very  probably  the trace of
the warm dust within the molecular/dusty  torus} invoked in the unified model of
AGN  since:  a) its  direction  is  perpendicular  to  the  axis  of  the  inner
ionization  cone and to the direction of the radio jet  originating  from S1; b)
its direction  follows  within  1\degr\ that of the small scale (1pc) radio disk
recently  discovered  by Gallimore et al.  (1997) with the VLBA  interpreted  by
these authors as the ionized outer envelope of the torus; c) the observed K flux
ratio of the core to the  region at r = 15 pc, is the same as the ratio of black
body emission at 1500 K and 600 K, respectively;  such a set of temperatures and
radius is consistent  with a simple model of grains  heated by a central  source
(T$_d^5  \propto  L_{AGN}  $r$^{-2}$),  the hot dust (1500 K) being located at a
radius of 1.5 pc and the warm dust (600 K) at a radius of 15 pc.

Concerning  the  elongated  S-shaped  feature  ($\pm$ 20 pc on each  side of the
unresolved  core), a first question  arises about the emission  mechanism at 2.2
$\mu $m.  We envisaged that the 2.2 $\mu $m emission is free-free radiation from
ionized gas stripped off the inner edges of the molecular/dusty torus and driven
away  along the radio  axis/inner  ionization  cone.  Under this  scenario,  the
related  radio  emission  at  6cm,  proportional  to the K  emission,  would  be
$\approx$ 300 mJy, a value in rough  agreement with that obtained by Ulvestad et
al.  (1987).  However, this interpretation would be in conflict with the finding
by Gallimore et al.  (1996a) that the radio emission along the jet (from S1 and
up to $\approx$ 0\farcs4 N) has an increasingly steep spectrum, fully consistent
with synchrotron emission.  We are left with the assumption that the emission at
2.2 $\mu $m is from stellar photospheres or from warm dust.  It is true that the
observed  [J-K] color, at  $r\approx 0 \farcs 2$, both along the ESE-WNW belt at
P.A.  = 102\degr and along the pseudo-bar  within the S-shaped  feature, remains
very  similar, in the range 3.4 to 3.6.  As we have seen above, the ESE-WNW belt
emission  at  2.2  $\mu$m  is  consistent  with  warm  dust  emission  (600K)  :
therefore, we might envisage that the 2.2 $\mu$m along the pseudo-bar within the
S-shaped feature originates from warm dust at the back of NLR clouds shaded from
the UV photons in the ionizing cone of the AGN.  As a very rough estimate, if we
assume a dust  temperature of 600K and a black-body  emission, then the measured
flux of 90 mJy in the  {0\farcs  2 $\times$  0 \farcs 3} area of the  pseudo-bar
would  correspond  to a  bolometric  luminosity  of the  dust  of 1.8  10$^{10}$
L$_\odot$,  i.e.  one tenth of the  bolometric  luminosity of NGC 1068.  A lower
temperature would lead to much larger luminosities that must be ruled out :  for
instance 6  10$^{13}$  L$_\odot$  if  T$_{dust}$  = 300K.  Starlight  is totally
unable to produce so large a dust temperature  \footnote{600K is the temperature
of grains at 4 10$^{-4}$ pc of a B2 star, a distance one thousand  times smaller
than  the  mean  distance  between  stars  if the  luminosity  of 1.8  10$^{10}$
L$_\odot$  was  accounted  for by a cluster  of B2 stars  within  the 0 \farcs 2
$\times$ 0 \farcs 3} and another dust heating mechanism should be invoked :  one
is direct  exposure to X-rays  emitted by the central  engine,  another could be
shocks, either through direct dust heating within the shock or through secondary
heating  by UV  photons  from  the  shocked  gas.  The  central  pseudo-bar  may
correspond  to dust in direct view of the AGN, but then the S-shape would not be
explained in this  scheme.  Shocks at  different  locations  may  overcome  this
difficulty ; in terms of energy balance, since models of fast, radiative  shocks
are indeed able to account for the total number of ionizing  photons produced in
a Seyfert  nucleus  (e.g.  Dopita \&  Sutherland,  1995), we can  assume  that a
significant  fraction  of this UV flux can  efficiently  heat the  dust.  On the
basis of this  sole data set, it is  difficult  to go beyond  this  stage in the
interpretation of the origin of the 2.2 $\mu $m photons.

Else, the  S-shape is  intriguing  and might tell us about the origin of the 2.2
$\mu $m emission.  As in the case of NGC 2110  (Mulchaey et al., 1994), we might
contemplate an interpretation involving the ejection of optical-emitting gas and
radio-emitting gas along the same axis although with different  velocities.  But
then, the emission  mechanism  at 2.2 $\mu $m would be  dominated  by  free-free
emission  which  seems to be  implausible.  We have  considered  an  alternative
scenario, involving the presence of a spiral and bar structure made of a mixture
of stellar and gaseous  components.  This later  scenario seems  promising as it
provides  as well clues on the central  engine  feeding.  Indeed, the bar within
bar  mechanism  has been  studied in details  through  N-body  simulations  of a
stellar and gaseous component  mixture  (Athanassoula,  1994).  As gas is pushed
inward  along the bar, it forms a  bar-unstable  disc on a smaller  scale, which
will  generate in turn a smaller  bar.  This  mechanism  could repeat  itself on
scales which are smaller and smaller.  The dynamical  behavior of such a system
has been modeled with various  configurations  (Combes, 1994b ; Friedli, 1996),
the prediction being that the innermost disc is in relation with the presence of
a bar about twice as large (Combes, private  communication).  In the case of NGC
1068, we noticed  that each of the radial  brightness  distributions  in J and H
(Fig.  \ref{F-raw}-f)  is  quite  well  fitted  by two  exponential  disks  with
characteristic radii of r$_{e}$ = 220 pc, r$_{i}$ = 19 pc.  Considering that the
J band is  mainly  a  tracer  of  stars,  then  the  scale-length  of the  inner
exponential  disk, 19 pc, is about what would be obtained  after a few rotations
of a 40 pc bar, the size of the elongated NS feature.  Under this  scenario, the
2.2 $\mu $m emission would be related as well to the stellar component, although
the presence of gas and dust taking part in the dynamics is not unlikely.

 Detailed  information on the kinematics of the innermost
regions are required to distinguish  between the two possibilities,  calling for
the obtainment of  2D-spectroscopic  data sets at high angular  resolution.  Yet,
the presence of two other  bar/spiral  systems on larger  scales in NGC 1068 and
the natural  explanation  for inward mass transfer, lead us to conclude that the
S-shaped  structure  may represent  the third level of an  interwoven  system of
bars, spiral and ring structures,  bringing material inward to build up and feed
the black-hole/accretion-disk system (Schlosman et al.  1989).  This constitutes
the first direct suggestion of the existence of a micro-barred/spiral  structure
in an AGN.

\acknowledgements   We  are  gratefully  indebted  to  Z.  Tsvetanov  for  making
available to us the fully  calibrated and aligned  HST/WFPC2  images and to J.P.
V\'eran for  reconstructing  PSFs with his powerful method.  We warmly thank the
CFHT team who allowed successful {\sl Pueo} observations. Thanks are extended to
the referee, J. Gallimore, who made several useful suggestions and to 
C. and G. Robichez for pertinent comments that helped improving the manuscript. 

 \end{document}